\documentclass[twocolumn]{svjour3}         
\usepackage{graphicx}
\usepackage{astron}
\usepackage{hyperref}                 

\begin{document}
\title{New $^{13}$CO observations towards the Keplerian gaseous disk around R Mon
}


\author{T. Alonso-Albi \and A. Fuente \and R. Bachiller \and A. Natta \and L. Testi \and R. Neri \and P. Planesas
}


\institute{
T. Alonso-Albi, A. Fuente, R. Bachiller, P. Planesas \at
Observatorio Astron\'{o}mico Nacional (OAN), Apdo 112, E-28800 Alcal\'{a} de Henares, Spain
\and
A. Natta, L. Testi \at
Observatorio Astrofisico di Arcetri, INAF, Largo E. Fermi 5, 50125 Firenze, Italy
\and
R. Neri \at
Institut de Radio Astronomie Milimetrique, 300 Rue de la Piscine, Domaine Universitaire de Grenoble, F-38406 St. Martin d'H\`{e}res, France
}

\date{Received: date / Accepted: date}

\maketitle

\begin{abstract}
We present new high angular resolution observations of the $^{13}$CO 1$\rightarrow$0 rotational line towards the HBe star R Mon, obtained with the IRAM Bure Interferometer, along with the previous results in the observed transitions $^{12}$CO 1$\rightarrow$0 and 2$\rightarrow$1. We have used a flat disk model to fit the $^{12}$CO 1$\rightarrow$0 and 2$\rightarrow$1, and $^{13}$CO 1$\rightarrow$0 emission, in a strip perpendicular to the outflow axis. The model assumes standard abundances (X($^{12}$CO) = 8$\cdot$10$^{-5}$, X($^{13}$CO) = 9$\cdot$10$^{-7}$), radial potential temperature and density laws (T (K) = T$_{o}$ r$^{-q}$, $\rho$~(cm$^{-3}$)~=~$\rho_{o}$ r$^{-p}$), and Local Thermodynamic Equilibrium. The $^{13}$CO and $^{12}$CO emission is consistent with a flat disk at an inclination angle of 20$^{\circ}$ in Keplerian rotation around the star. The gaseous disk is fitted with a mass of 0.014 M$_\odot$, an outer radius of 1500 AU, a temperature of 4500 K at the inner radius (1 AU), and values of q=0.62 and p=1.3 for the indexes of the temperature and density laws. This new $^{13}$CO observations allow us to conclude that the disk around R Mon is flat. Our result confirms previous works suggesting a predominant flat geometry in disks around early Be stars.


\end{abstract}

\section{Introduction}
\label{intro}
A big theoretical and observational effort has been done in recent years for the understanding of the disk occurrence and evolution in Herbig Ae/Be stars. Evidence of the existence of dusty and gaseous circumstellar disks around some HBe stars exists at optical, near and mid IR (see for example \cite*{mee01}, \cite*{vin02}, \cite*{mil01}, \cite*{ack05}, and more recently \cite*{ise06}). It is generally accepted the presence of disks similar to those in T Tauri stars for Herbig Ae and late (spectral type later than B7) Be stars, usually flared disks with inner rims, but the association of more massive early Be stars with disks is more uncertain, and the direct detection of a circumstellar disk around an HBe star at millimeter wavelengths has remained elusive until recently. The direct detection of a gaseous disk around R Mon also allows us to study the geometry and kinematics of a circumstellar disk in a B0 star.

R Mon is a very young B0 star with a T Tauri companion separated by 0.69". Because of its youth, R Mon is not directly visible in the optical, but appears as a resolved conical reflection nebula in scattered light. At infrared wavelengths, R Mon appears as a point source located 0.06"$\pm$0.02 south from the optical peak (\cite*{wei02}). Based on HI emission-line ratios, \cite{clo97} measured an extinction of Av = 13.1 mag towards the star that they interpreted as due to an optically thick disk of R = 100 AU.


\section{Previous results and new observations}
\label{prev}

\cite{fue03} detected a dusty disk around R Mon. New continuum observations (\cite*{fue06}, hereafter Paper I) carried out with the A configuration of the PdBI allowed us to constrain the size, mass, and spectral index of this circumstellar disk. Our results showed that the size of the dusty disk is 0.3", or 150 AU of radius, in agreement with the value derived by \cite{clo97}. Fitting the SED in the mm range (See \cite*{fue06} for details) we obtained our best fit with a dusty disk of mass M$_{d}~=~0.007 M_\odot$ and a spectral index $\beta$ = 0.3. The low value of $\beta$ suggested us that grain growth has proceeded to large sizes in the disk around R Mon. Typical values of 0.5-1 are found in Herbig Ae and T Tauri stars, also associated with grain growth.

Interferometric observations of $^{12}$CO 1$\rightarrow$0 and 2$\rightarrow$1 transitions were also presented in Paper I. We discussed our data by means of the results of a radiative transfer model applied to the observed molecular emission. We found that the observations were well fitted using a flat gaseous disk in Keplerian rotation, with an apparent inclination of 20 degrees from edge-on view, and a mass of 0.014 $M_\odot$. We were also able to constrain the mass of R Mon to 8$\pm$1 $M_\odot$. We found some keys to support a flat disk instead of a flared disk geometry. However, since the CO emission from this circumstellar disk is optically thick, we were not able to discard a flared geometry for the disk.

$^{13}$CO 1$\rightarrow$0 observations were carried out with the IRAM\footnote{IRAM is supported by INSU/CNRS (France), MPG (Germany), and IGN (Spain).} array at Plateau de Bure between January 2006 and March 2006, using AB configuration (beam 1.1"x0.7" at 110.2 GHz). The signal-to-noise ratio (5$\sigma$ detection) is not good enough to study the spatial distribution of this molecule, but allows us to conclude about the geometry of the disk around R Mon. All images in this paper are centered at RA 06:39:09.95, DEC +08:44:09.6 (J2000).

\begin{figure}
  \includegraphics[width=0.48\textwidth]{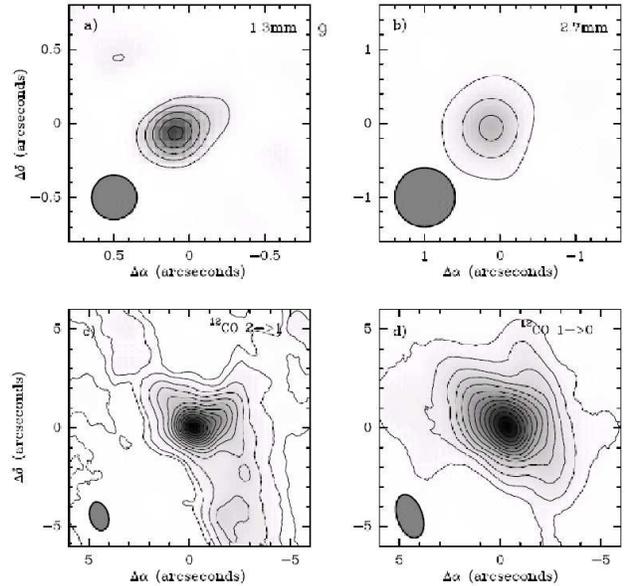}
\caption{(a) and (b) show the interferometric images of the circumstellar disk around R Mon in the continuum at 1.3mm (a) and 2.7mm (b). Contour levels are 1.5 mJy/beam to 8mJy/beam by 1.5mJy/beam in the 1.3mm image, and 0.75 mJy/beam to 8 mJy/beam by 1.5 mJy/beam in the 3mm image. The images have been obtained by convolving the clean components with a 0.3" circular beam in the case of the 1.3mm map, and a 0.8" circular beam in the case of the 2.7mm image. In (c) and (d) we show integrated intensity maps of the $^{12}$CO 2$\rightarrow$1 (c) and $^{12}$CO 1$\rightarrow$0 (d) rotational lines obtained with the PdBI. The contour levels are 0.25, 0.5 to 6 by steps of 0.5 Jy/beam*km/s in the $^{12}$CO 2$\rightarrow$1 image and 0.25, 0.5 to 6 by 0.25 Jy/beam*km/s in the $^{12}$CO 1$\rightarrow$0 image.}
\label{fig:1}       
\end{figure}

\begin{figure}
  \includegraphics[width=0.45\textwidth]{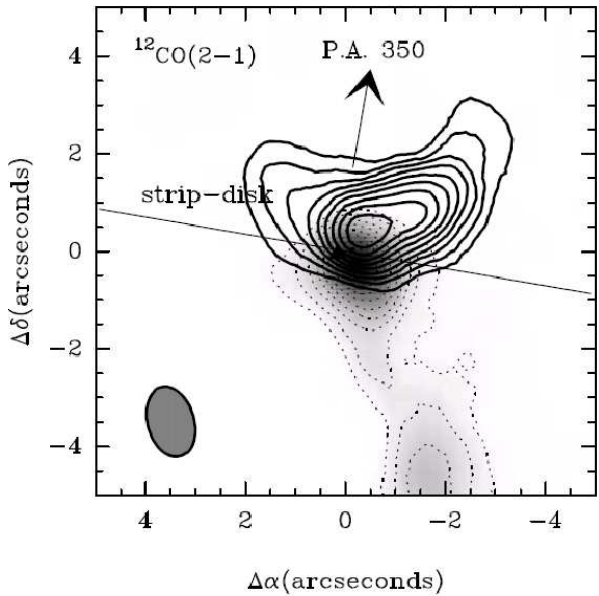}
\caption{This figure shows the integrated intensity map of the $^{12}$CO 2$\rightarrow$1 emission at blue (3.6-7 km/s) and red (12-15.4 km/s) velocities separately. The emission presents a bipolar morphology with a blue parabolic lobe towards the North and the red one towards the South. This N-S distribution is consistent with the morphology of the bipolar outflow at larger scales. The red lobe shows a jet-like morphology, i.e., we have detected a molecular jet associated with this B0 star. The strip-disk line shows the expected orientation of the disk.}
\label{fig:2}       
\end{figure}

\section{Model results and discussion}
The interferometric CO images (figure~\ref{fig:1}c and ~\ref{fig:1}d) show a compact molecular clump towards the star position with a size of 2.5" (2000 AU at d = 800 pc) in the $^{12}$CO 2$\rightarrow$1 line and a more extended and weaker component which follows the shape of the nebula. In figure~\ref{fig:2} the blueshifted and redshifted components are drawn separately. The orientation of the major axis of the disk is found to be perpendicular to the outflow axis.

%
\begin{table}
\centering
\begin{tabular}{ll}
\hline\noalign{\smallskip}
Parameter & Fitted value  \\
\noalign{\smallskip}\hline\noalign{\smallskip}
R Mon mass (M$_\odot$) & 8 $\pm$ 1 \\
i (deg) & 20 $\pm$ 5 \\
r$_o$ (AU) & 1 \\
T$_o$ (K) (r=r$_o$) & 4500 $\pm$ 500 \\
q & 0.62 $\pm$ 0.03 \\
Disk mass (M$_\odot$) & 0.014 $\pm$ 0.001 \\
p & 1.3 $\pm$ 0.1 \\
$\Delta v_{turb}$ (km/s) & 0.8 $\pm$ 0.2 \\
\noalign{\smallskip}\hline
\end{tabular}
\caption{R Mon flat disk parameters. Note errors are those obtained from the fitting procedure and do not account for other uncertainties inherent of this kind of calculations (instrumental errors, optical depth effects, etc.).
}
\label{tab:1}       

\end{table}

\begin{figure}
  \includegraphics[width=0.48\textwidth]{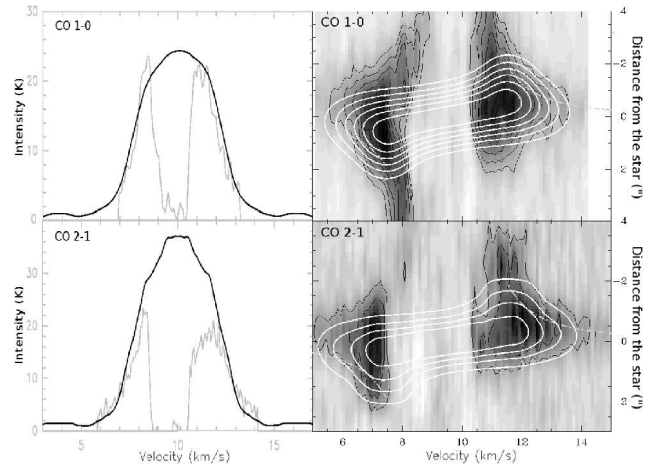}
\caption{At the right the PV diagrams of the CO rotational lines along the strip drawn in Fig.~\ref{fig:2} are shown. White contours superimposed show the synthesized P-V diagrams obtained with out flat disk model. Contour levels are 200-800 mJy/beam in steps of 100 mJy/beam (3.16 K) for the CO 1$\rightarrow$0 diagram, and 200-800 mJy/beam in steps of 200 mJy/beam (3.46 K) for the CO 2$\rightarrow$1 one. At the left, a comparison between the observed (grey) and the modeled (dark) spectra is shown for the star position. Note that the emission at the velocities of the molecular cloud (9.5 $\pm$ 2.5 km/s) has been filtered out by the interferometric observations.}
\label{fig:3}       
\end{figure}

The Position-Velocity (P-V) diagrams (figure~\ref{fig:3}) show the emission along the major axis of the disk. We have modeled the $^{12}$CO 2$\rightarrow$1 and 1$\rightarrow$0 emissions using a flat disk model. Our model assumes Local Thermodynamic Equilibrium (LTE) and standard $^{12}$CO abundance equal to $8\cdot$10$^{-5}$. With these assumptions, both, the CO 2$\rightarrow$1 and 1$\rightarrow$0 emissions are well fitted with a disk of 0.014 M$_\odot$ in Keplerian rotation around the star. The disk radius is 1500 AU and the inclination angle is 20$^{\circ}$ from an edge-on view. The mass of R Mon is fitted to 8 M$_\odot$. The gas kinetic temperature and density varies as T$~(K)~=~4500~r^{-q}$  and $\rho~$(cm$^{-3})~=~\rho_{o}$ r$^{-p}$, with q=0.62 and p=1.3. The density profile obtained with the model is consistent with the mass derived from the continuum data within a factor 2. A turbulent width component in the velocity profile is set to 0.8 km/s and added to the thermal width (table~\ref{tab:1}).

\begin{figure*}
\centering
  \includegraphics[scale=0.45]{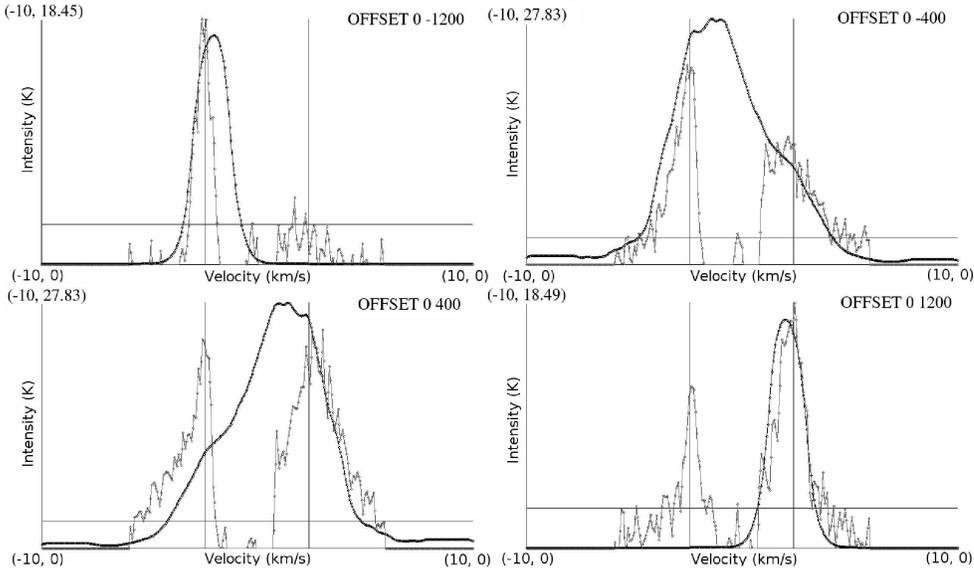}
\caption{These figures show the observed and the flat disk model predicted spectra over different positions along the major axis of the disk, for $^{12}$CO J = 2$\rightarrow$1. Offsets in AU are labelled at the top right corner in each figure. The horizontal and vertical lines represents respectively the assumed noise level (3 K) and interferometric filtering limits (v$_o$ $\pm$ 2.5 km/s,) in the fitting process.}
\label{fig:4}       
\end{figure*}

In figure~\ref{fig:4} the individual fit to some of the spectra (corresponding to certain position offset in AU along the major axis of the disk) for $^{12}$CO 2$\rightarrow$1 is shown. A very good fit is obtained all over the disk, even in other directions outside the major axis. This is the strongest evidence of a disk in Keplerian rotation obtained so far around an early Be star.

\begin{figure}
  \includegraphics[width=0.48\textwidth]{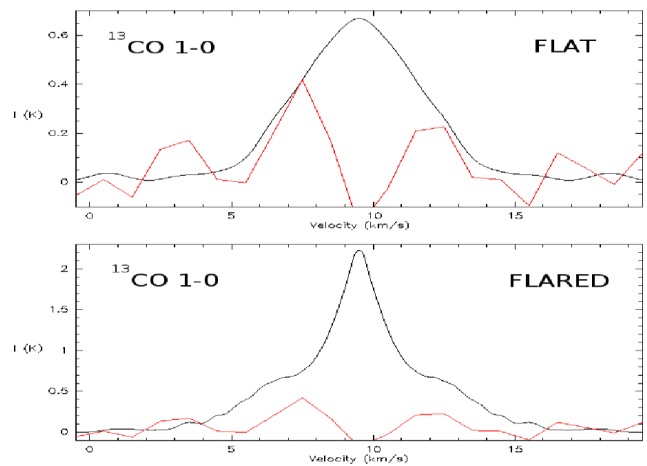}
\caption{The $^{13}$CO disk emission detected towards R Mon is well fitted by the flat disk model. The flared disk model (with a mass of 0.08 M$_\odot$) predicts an emission clearly above the observed values.}
\label{fig:5}       
\end{figure}

Since the $^{12}$CO lines are optically thick, only the surface of the disk is contributing to the emission, so the mass of the gaseous disk cannot be derived from these observations. As a consequence, the assumed geometry for the disk determines its mass. In fact, we reasonably fit the $^{12}$CO emission using both a flat and a flared disk model, but with a mass 8 times larger in the second case. We need to observe the rarer isotope $^{13}$CO to determine the opacity of the CO lines and derive the mass and density structure of the disk, and eventually the disk geometry (see discussion in Paper I). In figure~\ref{fig:5} we present the model results for $^{13}$CO, showing that the flat disk model also fits this emission using the same parameters and an abundance of 9$\cdot$10$^{-7}$ cm$^{-2}$ (90 times below $^{12}$CO standard abundance, as expected). The flared disk model predicts an emission that exceeds the observed one by a factor $\geq$2. As a result of these new observations and modellings, now we can confirm that the disk is flat. Furthermore, the $^{13}$CO/$^{12}$CO line intensity ratio reveals as a power diagnostic tool to discern between flared and flat disks.

\section{Conclusion}
Based on IR observations several authors have proposed that, contrary to HAe and TT stars, HBe stars seem to have geometrically flat disks (\cite*{ack05}). Our result in R Mon confirms this point. Moreover, the slope parameters for density and temperature distributions along the R Mon disk are similar to those found on TTs and some HAe (\cite*{dut04}), and, as most of them, is also in Keplerian rotation. The main difference respect to TTs and HAe stars is the mass of the disk, which is an order of magnitude lower in R Mon and in other HBe stars (\cite*{fue03}).

The similarities could indicate a common mechanism for star formation in the range $1-~8$ $M_\odot$. On the other hand, the different mass of the disks clearly indicates a shorter timescale for the evolution of disks around HBe stars, where the material is quickly dispersed. In fact, the low occurrence of disks found around HBe stars (\cite*{fue03})
supports this conclusion. A rapid evolution is also supported by the value of the spectral index $\beta$. This indicates the presence of large dust grains, which drops the optical depth, so the UV radiation penetrates to photoevaporate the external layers (\cite*{dul04}). Such disks could evolve to flat geometries in 10$^5$ years.

Future investigations will contribute to clarify this scenario. In this way, ALMA will be the definite instrument to study the properties of disks around intermediate mass protostars.

%
%
%

\bibliographystyle{astron}       

\end{document}